%Paper: hep-ph/9508286
%From: RAGAZZON@TRIESTE.INFN.IT
%Date: Fri, 11 Aug 1995 14:41:44 +0200 (WET-DST)

\pageno=0
\baselineskip=20pt plus 2pt
\magnification=\magstep1
\hsize=5.9truein
\vsize=8.7truein
\null
\centerline{ \bf MULTIPARTON INTERACTIONS AND PRODUCTION OF MINIJETS}
\centerline{ \bf IN HIGH ENERGY HADRONIC COLLISIONS}
\vskip .25in
\centerline{R. Ragazzon and D. Treleani}
\vskip .05in
\centerline{\it Dipartimento di Fisica Teorica dell'Universit\`a
and INFN Sezione di Trieste}
\centerline{\it Trieste, I 34014 Italy}
\vfill
\centerline{ABSTRACT}
\vskip .25in
\midinsert
\narrower\narrower
\noindent
We discuss the inclusive cross section to produce two minijets with
a large separation in rapidity in high energy hadronic collisions.
The contribution to the inclusive cross section from the
exchange of a BFKL Pomeron is compared with the contribution
from the exchange of two BFKL Pomerons, which is induced
by the unitarization of the semi-hard interaction. The effect of
the multiple exchange is studied both as a function of the azimuthal
correlation and as a function of the transverse momentum of the
observed minijets.
\endinsert
\vfill\eject
\par	{\bf 1. Introduction}
\vskip.25in
One of the main topics in perturbative QCD is presently
represented by semi-hard hadronic interactions, namely by hadron interactions
with momentum transfer constant with energy but large enough to apply
perturbation theory. One of the characteristic features of this kinematical
regime is the large size of the corresponding cross
sections, which, although in the perturbative domain, rise
rapidly with energy. In fact, already at the energies of present hadron
colliders,
one may easily obtain semi-hard cross sections whose size is comparable
to the total hadronic
cross section[1,2]. At the partonic level,
in a typical interaction configuration, one of the two interacting
partons has a finite fraction of
the parent's hadron momentum while the other one has a momentum fraction
close to zero. The separation in rapidity of the two partons is therefore
increasingly large with energy and, in the parton-parton c.m. system,
the transverse momentum
exchange is small with respect to the longitudinal momenta.
The Regge limit is then approached in semi-hard interactions
not only in the whole hadron-hadron process but also in the
underlying parton-parton interactions.
\par
When considering the large $p_t$ regime the momentum exchange is
of the order of the incoming partons momenta.
At the parton level such a large scale factor can be transferred
only in a few interaction vertices and, as a result,
the elastic two body parton collision is a good first
order approximation to the elementary
partonic interaction. In the semi-hard regime,  since the semi-hard scale is
small with respect to the
total energy available, there are several parton vertices with momentum
exchange
of the order of the semi-hard scale. A consequence is that
all semi-hard radiated gluons are to be taken explicitly into account
for a proper factorization of the semi-hard component of the interaction.
When the $2\to n$, rather than the $2\to2$,
is the parton subprocess relevant to
the semi-hard component of the hadronic interaction, a difficulty arises
in constructing an inclusive cross section, where only few of the radiated
partons are actually detected as mini-jets in the final state.
In fact one is not allowed any more
to use the lowest order tree diagram to represent
the parton amplitude, since the tree level amplitude is singular in the soft
and
collinear limit. To avoid the infrared problem one faces when evaluating
an inclusive cross section, one needs to keep virtual corrections
explicitly into account and, as a consequence,
the elementary subprocess acquires a non trivial
structure. The problem has been addressed already several years ago in a
series of papers by Lipatov and collaborators[3] . Lipatov's solution
is the BFKL Pomeron: the partonic interaction is described by the
exchange of a gluon ladder structure with vacuum quantum numbers
in the $t$-channel. The $s$-channel discontinuity of
the BFKL Pomeron represents the production of the semi-hard gluons.
In the limit in which the transverse momenta are always negligible
with respect to the longitudinal ones, the steps of the ladder are ordered in
rapidity
and dynamics is greatly simplified.
Indeed the simplified kinematics lets one to isolate
the two basic elements which build up the ladder:
\item{a-}
The gauge independent non-local vertices, which keep into account the dominant
term, in the
$t/s\to0$ limit, of the diagrams with
gluon emission from all near-by lines, and
\item{b-}
the Reggeization of the $t$-channel gluons, which is the virtual correction
that
allows a solution to the infrared problem.
\par\noindent
The ladder structure can be iterated in the $t$-channel, which may be expressed
as an integral equation, the Lipatov's equation. Lipatov's equation allows
an analytic solution free from infrared (and ultraviolet) singularities.
One obtains in this way an explicit expression for
the cross section where two gluons interact producing many gluons
and two of them, the ones nearby in rapidity to the interacting partons,
are observed. If $y$ is the separation in rapidity of
the interacting gluons and $k_a$, $k_b$ are the transverse momenta
of the observed ones, the inclusive cross section can be expressed as

$${d\hat{\sigma}\over d^2k_ad^2k_b}=\biggl[{C_A\alpha_s\over k_a^2}\biggr]
     f(k_a,k_b,y)\biggl[{C_A\alpha_s\over k_b^2}\biggr]\eqno(1)$$

\par\noindent
where $C_A=N_c$ is the number of colors, $\alpha_s$ is the strong coupling
constant and $f(k_a,k_b,y)$ is the inverse Laplace transform of the
solution to Lipatov's
equation. Actually:

$$f(k_a,k_b,y)={1\over(2\pi)^2k_ak_b}\sum_{{\rm n}=-\infty}^{+\infty}
   e^{i{\rm n}\phi}\int_{-\infty}^{+\infty}d\nu e^{\omega(\nu,{\rm n})y}
   e^{i\nu{\rm ln}(k_a^2/k_b^2)}\eqno(2)$$

\par\noindent
where $\phi$ is the azimuthal angle between the observed gluons,

$$\omega(\nu,{\rm n})=-2{\alpha_sN_c\over\pi}\Re
    \biggl[\psi\Bigl({\mid{\rm
n}\mid+1\over2}+i\nu\Bigr)-\psi(1)\biggr]\eqno(3)$$

\noindent
and

$$\psi(z)={d{\rm ln}\Gamma(z)\over dz}\eqno(4)$$

\noindent
is the Digamma function. The inclusive cross section for production
of two minijets, as a result of a BFKL Pomeron exchange, is obtained by folding
Eq.(1) with the structure
functions of the interacting hadrons $A$ and $B$:

$${d\sigma\over dx_Adx_Bd^2k_ad^2k_b}=
  f_{eff}(x_A,k_a^2)f_{eff}(x_B,k_b^2)
  {d\hat{\sigma}\over d^2k_ad^2k_b}\eqno(5)$$

\noindent
where $f_{eff}$ is the effective structure function

$$f_{eff}(x)=G(x)+{4\over 9}\sum_f\Bigl[Q_f(x)+{\bar Q}_f(x)\Bigr]
  \eqno(6)$$

\noindent
namely the gluon structure function plus $4/9$ of the quark
and anti-quark structure functions with flavour $f$.
The expression for the cross section in Eq.(1)
correlates the azimuthal angle $\phi$ with the distance in rapidity of the
observed partons.
The differential cross section in Eq.(1) may be easily integrated,
at $\phi$ fixed, on $k_a$ and $k_b$
down to the lower cut off $k_m$, which represents the threshold
in transverse momentum that
allows a parton to be observed as a minijet in the final state.
This yields:

$${d\hat{\sigma}\over d\phi}=
    {(C_A\alpha_s)^2\over2\pi}{1\over k_m^2}\sum_{{\rm n}=-\infty}^{+\infty}
    e^{i{\rm n}\phi}\int_{-\infty}^{+\infty}d\nu
  {e^{\omega(\nu,{\rm n})y}\over 1+4\nu^2}\eqno(7)$$

\noindent
which is a suitable expression to study the azimuthal correlation
of the observed partons as a function of the rapidity difference $y$.
Simple expressions may also be obtained for the cross section
where the momentum of one of the two observed gluons has been
integrated down to the lower limit $k_m$:

$${d\hat{\sigma}\over d^2 k_a}=
    {(C_A\alpha_s)^2\over2\pi}{1\over k_a^3k_m}
   \int_{-\infty}^{+\infty} d\nu
  {e^{\omega(\nu,0)y}\over 2i\nu+1}
  \biggl({k_a\over k_m}\biggr)^{2i\nu}\eqno(8)$$

\noindent
and for the cross section where both observed gluons momenta have
been integrated down to $k_m$:

$$\hat{\sigma}=
    {(C_A\alpha_s)^2\over k_m^2}
   \int_{-\infty}^{+\infty} d\nu
  {e^{\omega(\nu,0)y}\over 1+4\nu^2}
  \eqno(9)$$

\noindent
The high energy behaviour of the integrated cross section is estimated by
evaluating the asymptotic limit of the integral in Eq.(9) for large $y$[4]:

$$\hat{\sigma}\to
 {\pi(C_A\alpha_s)^2\over2k_m^2}
 {exp\bigl[4{\rm ln}2N_c\alpha_sy/\pi\bigr]\over
 \bigl[7\zeta(3)N_c\alpha_sy/2\bigr]^{1/2}}\eqno(10)$$

\noindent
where $\zeta$ is the Riemann zeta function.
Eq.(10) shows a
rapid growth, roughly linear with the parton-parton c.m. energy, of
the partonic cross section corresponding to the exchange of a BFKL Pomeron
and gives a justification to the large size of the semi-hard cross section.
\par
The possibility to describe the elementary parton process by means
of Lipatov's dynamics has been considered recently in a series of
papers[5].
One of the main points of interest is the search for clear signatures of the
underlying parton dynamics
in the final state of high energy hadronic collisions. Correlations in
transverse momentum
and azimuthal angle, as a function of the distance in rapidity $y$
of final state minijets, have been therefore estimated according
to the expectations of the Lipatov's picture of the interaction as
expressed by Eq.(7) and (8)[6]. On the other hand, to approach Lipatov's limit,
one needs to keep the lower threshold of the transverse momenta
of the observed minijets as small as
possible, compatibly with the requirement of being still in the perturbative
regime.
As shown by Eq.(10) smaller values of $k_m$ correspond to larger values
of $\hat{\sigma}$. As a consequence of the larger probability of the elementary
partonic intercourse one is therefore forced to take into account
the possibility of having several elementary partonic collisions in each
inelastic hadronic event, in order to implement unitarity.
\par
In the present paper, by assuming the validity of the AGK cutting rules
in semi-hard interactions, we unitarize the semi-hard cross section
and we derive the most general correction term to the inclusive cross section
in Eq.(5). The correlations among the minijets observed in the final state
are then estimated, considering the simplest possibility of multiple parton
interaction, and are
compared with the expectation from the single BFKL Pomeron exchange.
The paper is organized as follows: in the next section the unitarity
correction to the single scattering term is derived. In the following
paragraph a few numerical estimates are presented, with the purpose of
giving an indication on the kinematical region where corrections
can be expected to become sizeable. In the last section the general features
of the unitarization of the single scattering term
are summarized and a few general conclusions are drawn.
\vskip .25in
\par    {\bf 2. General framework for muliparton interactions}
\vskip .15in
\par
In order to approach the problem of multiparton interactions,
with the purpose of obtaining for the inclusive
cross section an expression which is more general with respect to
Eq.(5),
we find it appropriate to introduce a functional
formalism, which keeps to a minimum level the occurrence of cumbersome
expressions. As a preliminary step, we show how to derive all the relevant
inclusive cross sections in the simple case of a single parton-parton
collision with fixed  fractional longitudinal momenta $(x,x')$.
Let us introduce the functional

$$\hat{\Theta}[x,x';z]\equiv\sum_n\int{d\hat{\sigma}_n(x,x')\over dk_1\dots
dk_n}z(k_1)
\dots z(k_n)dk_1\dots dk_n \eqno(11)$$

\noindent
where  $z$ is the argument of the functional and $d\hat{\sigma}_n$ is the
differential cross section to produce $n$ partons with momenta
$(k_1,\dots,k_n)$. Obviously the value of the functional
for $z=1$ is the semi-hard parton cross section
$\hat{\sigma}(x,x')$. Actually

$$\hat{\Theta}[x,x';1]=\hat{\sigma}(x,x')\eqno(12)$$

\noindent
All the inclusive cross sections are generated
by taking an appropriate number of functional derivatives of the
generating functional with respect
to $z$[7]:

$${d\hat{\sigma} (x,x')^{incl}\over dk_1\dots dk_n}=
{\delta\hat{\Theta} [x,x';z]\over\delta z(k_1)
\dots\delta z(k_n)}\biggm|_{z=1}.\eqno(13)$$

\par
To obtain the inclusive cross section in the case of
the actual hadronic collision
a more elaborate analysis is needed. In the case of soft
interactions
multi-Reggeon exchanges are conveniently taken into account by making use of
the AGK cutting rules[8]. Although no general
proof of their validity is available in the case of
semi-hard interactions, it has nevertheless been possible to show that
the cutting rules hold for one of the components of the
interaction which is leading in the large-$\hat{s}$
fixed-$\hat{t}$ limit[9]. If one assumes the validity of
the cutting rules for semi-hard interactions, one is allowed to represent the
semi-hard cross section $\sigma_H$ as a probabilistic distribution of
multiple semi-hard parton collisions[10].
The most general expression for
$\sigma_H$ requires however the introduction of
the whole infinite set of multiparton distributions[11], which keep into
account hadron fluctuations in the parton number:

$$\sigma_H=\int d^2\beta\sigma_H(\beta)$$

$$\eqalign{\sigma_H(\beta)=\int&\sum_n\sum_m{1\over n!}
W_A^{(n)}(u_1\dots u_n)
{1\over m!}W_B^{(m)}(u_1'-\beta\dots u_m'-\beta)\cr
\times&\Bigl\{1-\prod_{i=1}^n\prod_{j=1}^m\bigl[1-\hat{\sigma}(u_i,u'_j)\bigr]
   \Bigr\}\prod dudu'}
\eqno(14)$$

\noindent
Here the $W^{(k)}(u_1\dots u_k)$ are the exclusive $k$-body parton
distribution,
namely the probabilities to find a hadron in a fluctuation with $k$ partons
with
coordinates $u_1\dots u_k$,
 $u_i\equiv(b_i,x_i)$ standing for the transverse partonic coordinate $(b_i)$
and longitudinal fractional momentum $(x_i)$.
$\beta$ is the impact parameter between the two interacting hadrons
and $\hat{\sigma}(u_i,u'_j)$,
 represents
the probability for the parton $i$
of the $A$-hadron to have an hard interaction with the parton $j$
of the $B$-hadron.
The semi-hard cross section is constructed by summing over all possible
partonic configurations of the two interacting hadrons (the sums over
$n$ and $m$) and, for each configuration with $n$ $A$-partons and
$m$ $B$-partons, summing over all possible multiple partonic
interactions. This last sum is constructed by asking for the
probability of no interaction between the two configurations
(actually $\prod_{i=1}^n\prod_{j=1}^m[1-\hat{\sigma}_{i,j}]$ ). The
difference from one of the probability of no interaction
gives the sum over all
semi-hard interactions. $\sigma_H(\beta)$ is then the probability to have
at least one semihard parton interaction when the impact parameter in
the hadronic collision is equal to $\beta$. The semi-hard cross section is
obtained by integrating the probability $\sigma_H(\beta)$ on the impact
parameter.
Analogously, the elementary semi-hard cross section $\hat{\sigma}(x,x')$ is
obtained by integrating the elementary interaction probability
$\hat{\sigma}(u,u')$
on the relative transverse coordinate ${\bf b}-{\bf b}'$.
\noindent
The expansion of
$\sigma_H(\beta)$ as a sum on multiple interactions reads:

$$\eqalign{\sigma_H(\beta)=
\int&\sum_n\sum_m{1\over n!}W_A^{(n)}(u_1\dots u_n){1\over m!}
W_B^{(m)}(u_1'-\beta\dots u_m'-\beta)\cr
\times&{\cal S}\sum_{N=1}^Q
  {Q\choose N}\hat{\sigma}_1\dots\hat{\sigma}_{N}
  (1-\hat{\sigma}_{N+1})\dots(1-\hat{\sigma}_Q)}\eqno(15)$$

\noindent
${\cal S}$ is a symmetrizing operator, which one may conveniently introduce
taking advantage of the symmetry of $W^{(k)}$ for permutations of
the arguments[12], and the index $N$
counts the interactions which, for a given configuration with
$n$ A-partons and $m$ B-partons, range in number from 1 to  $Q=nm$.
As a matter of fact, the main advantage of
Eq.(15) is the clear separation between real and virtual contributions
to the semihard cross section. More precisely, after summing, according with
the AGK cutting rules, over
all discontinuities of the semi-hard amplitudes, which contribute to the
inelastic process of interest,
the product $\hat{\sigma}_1\dots \hat{\sigma}_N$ is the remnant of the
contribution from the
real production terms. The product
$(1-\hat{\sigma}_{N+1})\dots (1-\hat{\sigma}_Q)$ is, on the contrary, the
remnant of
the contribution of the virtual corrections[13].
The replacement $\hat{\sigma}_k\rightarrow\hat{\Theta}_k[z]$ in the former
product,
corresponding to the real production process, allows one to
generalize the functional in Eq.(11) and to obtain the
inclusive cross sections in the most general case of multiple
parton interactions. One may therefore write

$$\eqalign{\Theta_H[\beta;z]=
\int&\sum_n\sum_m{1\over n!}W_A^{(n)}(u_1\dots u_n){1\over m!}
W_B^{(m)}(u_1'\dots u_m')\cr
\times&{\cal S}\sum_{N=1}^Q
  {Q\choose N}\hat{\Theta}_1[z]\dots\hat{\Theta}_{N}[z]
  (1-\hat{\sigma}_{N+1})\dots(1-\hat{\sigma}_Q)\prod dudu'}\eqno(16)$$

\noindent
which gives the required inclusive cross sections
via the relation

$${d\sigma_H^{incl}\over dk_1\dots dk_n}=\int d^2\beta{\delta\Theta_H[\beta;z]
\over\delta z(k_1)
\dots\delta z(k_n)}\biggm|_{z=1}.\eqno(17)$$

\noindent
For later convenience
$\Theta_H[\beta;z]$ can also be expressed as

$$\eqalign{\Theta_H [\beta;z]=
\int\sum_n&\sum_m{1\over n!}W_A^{(n)}(u_1\dots u_n){1\over m!}
W_B^{(m)}(u_1'-\beta\dots u_m'-\beta)\cr
\times\Bigl\{&\prod_{i=1}^n\prod_{j=1}^m\bigl[1+\hat{\Theta}[u_i,u'_j;z]
-\hat{\sigma}(u_i,u'_j)\bigr] \cr
 -&\prod_{i=1}^n\prod_{j=1}^m\bigl [ 1-
\hat{\sigma}(u_i,u'_j)\bigr]\Bigr\}
\prod dudu'}\eqno(18)$$

\par
We are now in a position to discuss the processes we are interested
in, namely the events in which only two mini-jets are tagged.
By setting $n=2$ in Eq. (17) and using the second expression
for $\Theta_H[\beta;z]$, a lengthy but simple algebra yields

$$\eqalign{{d\sigma_H^{incl}(\beta)\over dk_1 dk_2}=&\int D_A^{(1)}(u)
D_B^{(1)}(u'-\beta){d\hat{\sigma}(u,u')\over dk_1dk_2}dudu'\cr
+&\int D_A^{(2)}(u,v)D_B^{(2)}(u'-\beta,v'-\beta)
{d\hat{\sigma}(u,u')\over dk_1}{d\hat{\sigma}(v,v')\over dk_2}dudu'dvdv'\cr
+&\int D_A^{(1)}(u)D_B^{(2)}(u'-\beta,v'-\beta)
{d\hat{\sigma}(u,u')\over dk_1}{d\hat{\sigma}(u,v')\over dk_2}dudu'dv'\cr
+&\int D_A^{(2)}(u,v)D_B^{(1)}(u'-\beta)
{d\hat{\sigma}(u,u')\over dk_1}{d\hat{\sigma}(v,u')\over dk_2}dudvdu'}
\eqno(19)$$

\noindent
where $D^{(1)}(u)$ and $D^{(2)}(u,v)$ are the one-body and two-body
inclusive distributions [10]:

$$\eqalign{D^{(1)}(u)=W^{(1)}(&u)+\int W^{(2)}(u,u')du'+{1\over 2}
		    \int W^{(3)}(u,u',u'')du'du''+\dots\cr
	     D^{(2)}(u_1,u_2)=W^{(2)}(&u_1,u_2)+\int W^{(3)}(u_1,u_2,u')du'\cr
		    &+{1\over 2}
		    \int W^{(4)}(u_1,u_2,u',u'')du'du''\dots\cr
	}\eqno(20)$$

\noindent
In the r.h.s. of Eq. (19) every term has a clear physical interpretation.
The first convolution is nothing but the usual
single-collision contribution to the semi-hard cross section.
The second term corresponds
to two disconnected partonic collisions; finally, the last two entries
correspond to those events in which a parton from hadron $A$ or $B$ has
suffered a rescattering on hadron $B$ or $A$ respectively.
\par
{}From ref.[14] we know that the average number of rescatterings
can be safely neglected in a typical hadron-hadron collision and
for values of $k_m$ which allow the final state parton to be observed as an
actual
minijet in the final state. We are therefore allowed to neglect the last two
terms in the r.h.s.
of Eq. (19).
The two-body inclusive distribution $D^{(2)}$ may be expressed
by introducing the two body parton correlation $C^{(2)}$:

$$D^{(2)}(u_1,u_2)\equiv D^{(1)}(u_1)D^{(1)}(u_2)+{1\over 2}
C^{(2)}(u_1,u_2)\eqno(21)$$

\noindent
If one neglects both rescatterings and correlations in Eq.(19),
one is left with the following simplified expression for the inclusive
cross section:

$$\eqalign{{d\sigma_H^{incl}\over dk_1dk_2}&=
\int d^2\beta \Bigl[
 D_A^{(1)}\otimes {d\hat{\sigma}\over dk_1dk_2}\otimes D_B^{(1)}\cr
&+\bigl(D_A^{(1)}\otimes{d\hat{\sigma}\over dk_1}\otimes D_B^{(1)}\bigr)
\bigl(D_A^{(1)}\otimes{d\hat{\sigma}\over dk_2}\otimes D_B^{(1)}\bigr)\Bigr]}
\eqno(22)$$

\noindent
where $\otimes$ is a compact notation for the convolutions appearing
in Eq. (19). A possible further simplification follows from the assumption that
$D^{(1)}(u)$ has the factorized form

$$D^{(1)}(x,b)=f_{eff}(x)F({\bf b})\eqno(23)$$

\noindent
with the obvious normalizing condition

$$\int d^2 b F({\bf b})=1\eqno(24)$$

\noindent
By substituting Eq. (23) in Eq. (22) one obtains

$${d\sigma_H^{incl}\over dk_1dk_2}={d\sigma_s\over dk_1dk_2}+
{1\over \sigma_{eff}}{d\sigma_s\over dk_1}{d\sigma_s\over dk_2}
\eqno(25)$$

\noindent
where

$${1\over\sigma_{eff}}
\equiv\int d^2\beta\Bigl[\int d^2bF({\bf b})F({\bf b}-{\bf \beta})
\Bigr]^2\eqno(26)$$

\noindent
and $d\sigma_s$ is the single collision expression, obtained by
convoluting the elementary cross section with the usual one-body
parton distribution $f_{eff}(x)$.
\vskip.25in
\par    {\bf 3. Numerical estimates}
\vskip .15in
\par
The formalism described in the previous section is a rather general approach
to the problem of unitarity corrections in semi-hard interactions.
Indeed the expression for the inclusive cross section in Eq.(19) is
completely general in the probabilistic picture of the semi-hard
hadronic interaction. It is an exact consequence of the cross section
as expressed in Eq.(14), which finds its justification in the AGK cutting
rules[8]. In the inclusive cross section given by Eq.(19)
all possible multiple parton collisions are kept into account and
multiparton correlations are treated at all orders.
Consistently with the general principles, namely with the AGK
cancellation[8,15], the
double inclusive cross section depends only on the single and
double scattering terms.
For a quantitative estimate of the role of unitarity corrections
to the single scattering term, the required non perturbative input
is represented both by the one-body parton distribution $D^{(1)}$ and
by the two-body parton distribution $D^{(2)}$. The two-body parton
distribution contains an independent information on the hadron structure with
respect to $D^{(1)}$, actually the two-body parton correlation $C^{(2)}$.
While no experimental information is presently available
on $C^{(2)}$ an indication is available from CDF on the scale factor
$\sigma_{eff}$ which characterizes the double parton interactions[16].
We will therefore limit our numerical analysis to the simplified case where
$C^{(2)}$ is
neglected and only disconnected parton collisions are taken into account, in
such
a way that the inclusive cross section is expressed by Eq.(25).
All unitarity corrections to single scattering
are therefore expressed by the second term in Eq.(25), which is obtained
with the same input needed to evaluate the single scattering term, apart from
the scale factor $\sigma_{eff}$, that summarizes
all the geometrical details which enter in the unitarity correction.
\par
A few qualitative considerations are appropriate before illustrating the
results of
a quantitative analysis.
By introducing the jet rapidities $(y_a,y_b)$ and integrating in the transverse
momenta down to the lower
cut off $k_m$, while keeping fixed the azimuthal angle between the
observed minijets $\phi$, the inclusive cross section is expressed as:

$${d\sigma_H^{incl}\over d\phi dy_ady_b}={d\sigma_s\over d\phi dy_ady_b}+
{1\over \sigma_{eff}}{1\over 2\pi}{d\sigma_s\over dy_a}{d\sigma_s\over dy_b}
\eqno(27)$$

\noindent
In the limit of small relative rapidities $y=y_{a}-y_{b}$, a parton-parton
interaction produces only two final state partons. Since they are back-to
-back in $\phi$, the single collision expression $d\sigma_{s}/
 d\phi dy_{a}dy_{b}$
 is proportional to a Dirac delta $\delta (\phi-\pi)$.
This can be easily verified by setting $y=0$ in Eq.(7).
On the opposite side, that is, for large values of $y$, the leading
contribution to the r.h.s. of Eq.(7) comes from the $n=0$ term, for
which the partons are decorrelated in $\phi$. Physically, this is due
to the large number of gluons radiated  in the parton-parton interaction.
Indeed, the flattening of the $\phi$ distribution with increasing dijet
rapidity gap was suggested[6] as a signature for the BFKL
dynamics. From this point of view, a multiple partonic collision represents
a background process which mimics the effect of   multigluon emission.
In the r.h.s. of Eq. (27), this background is described by the term weighted
by the scale factor $1/\sigma_{eff}$.
The experimental indication on the scale factor is $5.4<\sigma_{eff}<29mb$
(90{\%}  C.L.)[16]. Unfortunately $\sigma_{eff}$ is not the only input variable
which is
still rather uncertain for a numerical computation. Indeed there is a large
ambiguity already to compute the
single scattering term. In fact to obtain the Lipatov's solution one needs to
neglect the running of the strong coupling constant, in such a way that
$\alpha_s$ has to be considered as a parameter in the actual evaluation
of $\hat{\sigma}$. Since the dependence
of $\hat{\sigma}$ on $\alpha_s$, as it may be seen in Eq.(10), is exponential
a numerical comparison of the two terms in Eq.(27) is rather uncertain.
\par\noindent
To have a quantitiative feeling of the importance of the unitarity
correction we have tried to estabilish a possible
sensible choice of the input values of $\sigma_{eff}$ and $\alpha_s$ by
making a comparison with available experimental data.
The experimental points in
fig.1 are the values of the cross section for production of minijets with
$k_m\ge5GeV$ measured
by UA1[2]. The dashed curves refer to the single scattering integrated
cross section with $\alpha_s=.34$
(upper curve) and $\alpha_s=.29$ (lower curve), corresponding to the values of
the running coupling constant at the scale $\sqrt{Q^2}=k_m/3$
and $\sqrt{Q^2}=k_m/2$ respectively. The structure functions are the HMRS(B)
structure functions[17]. The unitarized expression for the semihard cross
section
has a simple analytical representation when semi-hard rescatterings and
multi-parton correlations
are neglected[10]. Actually:

$$\sigma_H=\int d^2\beta\biggl(1-exp\Bigl[\sigma_s
 \int F_A({\bf b})F_B({\bf b}-\beta)d^2b\Bigr]\biggr)\eqno(28)$$

\noindent
where $\sigma_s$ is the integrated single scattering
inclusive cross section. The continuous curves
in fig.1 refer to the unitarized cross section
$\sigma_H$, as expressed in Eq.(28). For $F(b)$ we have taken
a gaussian, the width corresponding to a value of $\sigma_{eff}=20mb$.
The two curves refer to the two different choices of $\alpha_s$
mentioned above.
The region identified by the two continuous lines contains the
experimental points and
therefore gives an indication on possible meaningful input parameters.
One may also observe in fig.1 how the rise of the experimental cross section is
much
closer to the rise of the unitarized curves than to the rise of the single
scattering
term alone.
\par
Before moving to different values of energy it is worthy to briefly comment on
 $k_{m}$, which, to
some extent, is a free parameter.
 A low value of  $k_{m}$ corresponds to
semi-hard cross sections that are well above $\sigma_{eff}$ (in
the single collision approximation). In this
conditions the contribution from multiple scatterings is largely
dominant and the $\phi$ distribution is practically flat.
On the contrary, large values of $k_m$ correspond to semi-hard
cross sections that are negligible with respect to $\sigma_{eff}$ and
no unitarity correction is required. Keeping this in mind, we realize
that the interesting
 values of $k_{m}$ are those for which the total semi-hard
cross section is comparable to $\sigma_{eff}$. This criterion yields
$k_{m} \simeq 5.2\div6.1GeV$ at $\sqrt{s}=1.8 TeV$
and $k_{m}\simeq 11.2\div12.7 GeV$ at $\sqrt{s}=18 TeV$,
depending on the two different choices of values for $\alpha_s$ which we have
considered and for $\sigma_{eff}=20mb$.
\par
In order to have some quantitative indication on
the effect that unitarization produces on the expectations based on the BFKL
dynamics,
we have studied the azimuthal correlation of the observed minijets,
which, according with the BFKL dynamics has a distinctive dependence on the
distance in rapidity.
In fig. (2-a,b) we have plotted the differential cross
section Eq.(27) as a function of $\phi$, for fixed rapidities $(y_{a}
,y_{b})$ at $\sqrt{s}=1.8TeV$ (a) and $\sqrt{s}=18TeV$ (b)
(the normalization is such that the curves take a value equal to unity
at $\phi=0$ and $\phi=2\pi$).
The naive $\phi$ distribution, obtained by considering one elementary
interaction only, is represented by  the dashed  line, while the continuous
line
describes
the corrected distribution which takes into account an
arbitrary number of parton-parton collisions. The flattening caused by
the unitarity corrections is clearly visible: at $\sqrt{s}=1.8 TeV$,
 fig. (2-a),
the height of the central peak at $\phi=\pi$ is reduced by a factor
three approximately; the same trend, but with a stronger suppression of
the correlation, is present at higher energies, see fig. (2-b).
Figure (3) shows how the effect of unitarity corrections depends on the
cutoff $k_{m}$. By lowering this threshold we increase the semi-hard
cross section and, accordingly, we enhance the probability of having
several elementary parton collisions in each inelastic hadronic event.
As a consequence, we expect the tagged minijets to
 become less and less correlated
in the azimuthal angle $\phi$. This is confirmed by our plot which
 corresponds to
$\sqrt{s}=1.8 TeV$ and to a rapidity gap $y=5$, actually $y_{a}=2.5$ and
$y_{b}=-2.5$. The different choices of the cut off $k_m$ are
$k_m=7GeV$ (solid line), $k_m=6GeV$ (dashed line) and $K_m=5GeV$ (dotted
line).
It is worthwhile to stress that,
for the lower choice $k_{m}=5GeV$, we cannot distinguish the
$\phi$ distribution from a uniform one, unless
we perform a quite accurate measure at the 3{\%} level.
Finally,  fig. (4) shows how the correlation in the azimuthal angle
of the tagging jets fades away as the rapidity interval is increased.
\vskip.25in
\par	{\bf 4. Conclusions}
\vskip.15in
\par	Minijet physics is the ideal tool to study BFKL dynamics. Indeed
one comes closer and closer to the BFKL limiting case by keeping the
lower threshold in
transverse momentum $k_m$ of the observed minijets as small as possible.
However the region of small $k_m$ is also the region where unitarity
corrections
become increasingly important. In the present paper we have made an attempt
to estimate the unitarity corrections to the inclusive cross section for
producing two minijets.
After assuming the validity of the AGK cutting rules
in semihard interactions, we have kept into account unitarity corrections
by representing the hadronic process as
a probabilistic superposition of multiple BFKL Pomeron exchanges.
In the case of the inclusive cross section for producing two minijets,
only the single and the double scattering terms contribute.
With the purpose of making a quantitative estimate, we have considered the
simplest possibility
for the double scattering contribution. Actually we have neglected
semi-hard parton rescatterings in the interaction, and two-body parton
correlations in the two-body inclusive distributions. In this simplified case
the unitarity
correction depends on one single parameter only, namely $\sigma_{eff}$,
that is the scale factor one needs to introduce in order to obtain
the probability of the double interaction.
For a quantitative illustration of the effect of the
correction term, a second parameter which has to be fixed is the
strong coupling constant $\alpha_s$, whose value is not determined by the
BFKL dynamics.
Keeping into account the experimental
suggestion on $\sigma_{eff}$[16], we have fixed the input parameters by
comparing with the UA1 measurement of the semihard cross section
for production of minijets[2].
Having selected in this way a possible range of values for the parameters,
the indication we obtain from our numerical estimate is that at Tevatron energy
the correction term
to the single BFKL Pomeron exchange, depending on the actual quantity one is
considering,
may be larger than $100\%$ for
minijets with $k_m\simeq6GeV$.
When moving at LHC energies the same correction applies with
values of $k_m\simeq12 GeV$.
It is worthwhile noticing that, at Tevatron energy and with
$k_m\simeq6GeV$ the average invariant mass of a partonic
interaction is $\simeq.2TeV$, while at
LHC energies and with
$k_m\simeq12 GeV$, the average
invariant mass is $\simeq1TeV$. The expectation is therefore
that a secondary BFKL Pomeron is exchanged in a
large fraction of parton interactions at those values of invariant
mass and at the corresponding hadron-hadron c.m. energy.
A detailed experimental analysis of minijet production at Tevatron
would therefore be of
great importance both as a test of the BFKL approach,
and to access the non perturbative information on the hadron structure
which enters in the multiple parton interactions, whose
detailed knowledge is of growing importance
to understand hadron dynamics at higher energies.
\vskip.25in
\par	{\bf Acknowledgements}
\vskip.15in
\par\noindent
Helpful discussions with G. Calucci are gratefully acknowledged. This work
has been partially supported by the Italian Ministry of University and of
Scientific
and Technological Research with the Fondi per la Ricerca
Scientifica-Universit\`a
di Trieste.
\vfill
\eject
\par    {\bf References}
\vskip .15in
\item{1.} G. Pancheri and Y. Srivastava {\it Phys. Lett.} {\bf B182},
199 (1986); S. Lomatch, F.I. Olness and J.C. Collins {\it Nucl. Phys.}
{\bf B317}, 617 (1989).
\item{2.} C. Albajar et al. {\it Nucl. Phys.} {\bf B309},
405 (1988).
\item{3.} L.N. Lipatov, {\it Yad. Fiz.} {\bf 23}, 642 (1976);
E.A. Kuraev, L.N. Lipatov and V.S. Fadin, {\it Zh. Eksp. Teor. Fiz.}
{\bf 71}, 840 (1976) [{\it Sov. Phys. JEPT} {\bf 44}, 443 (1976)];
{\bf 72}, 377 (1977) [{\it Sov. Phys. JEPT} {\bf 45}, 199 (1977)];
Ya.Ya. Balitskii and L.N. Lipatov, {\it Yad. Fiz.} {\bf 28},
1597 (1978) [{\it Sov. J. Nucl. Phys.} {\bf 28}, 822 (1978)];
{\it Pis'ma Zh. Eksp. Teor. Fiz.} {\bf 30}, 383 (1979)
[{\it JETP Lett.} {\bf 30}, 355 (1979)].
\item{4.} A.H. Mueller and H. Navelet, {\it Nucl. Phys.}
{\bf B282}, 727 (1987).
\item{5.} N.N. Nikolaev, B.G. Zakharov and V.R. Zoller,
{\it J. Exp. Theor. Phys.} {\bf 78}, 806 (1994);
{\it Phys. Lett.} {\bf B328}, 486 (1994);
{\it JETP Lett.} {\bf 59}, 6 (1994);
Del Duca and C. R. Schmidt, preprint DESY-94-163, Oct 1994;
V. Del Duca, preprint DESY-95-023, Feb 1995;
V. Del Duca and C. R. Schmidt, {\it Phys. Rev.} {\bf D51},
2150 (1995).
\item{6.} V. Del Duca and C. R. Schmidt, {\it Phys. Rev.} {\bf D49},
4510 (1994); W.J. Stirling, {\it Nucl. Phys.} {\bf B423},
56 (1994).
\item{7.} I. V. Andreev, {\it Sov. J. Nucl. Phys.} {\bf 22},
92 (1976) [{\it Yad. Fiz.} {\bf 22}, 186 (1975)].
\item{8.} V. Abramovskii, V.N. Gribov and O.V. Kancheli, {\it Yad. Fiz.}
{\bf 18}, 595 (1973) [{\it Sov. J. Nucl. Phys.} {\bf 18}, 308 (1974) ];
I.G. Halliday and C.T. Sacharajda, {\it Phys. Rev.} {\bf D8}, 3598 (1973);
J. Koplik and A.H. Mueller, {\it Phys. Lett.} {\bf 58B}, 166 (1975);
L.D. Mc Lerran and J.H. Weiss, {\it Nucl. Phys.} {\bf B100}, 329 (1975).
\item{9.} G. Calucci and D. Treleani {\it Phys. Rev.} {\bf D49}, 138 (1994);
{\bf D50}, 4703 (1994); J. Bartels and M. W\"{u}sthoff, {\it Z. Phys.}
{\bf C66}, 157 (1995).
\item{10.} G. Calucci and D. Treleani, {\it Nucl. Phys.}
{\bf B} (Proc. Suppl.) 18C, 187 (1990) and
{\it Int. J. Mod. Phys.} {\bf A6}, 4375 (1991).
\item{11.} H.D. Politzer, {\it Nucl. Phys.} {\bf B172}, 349 (1980);
R.K. Ellis, R. Petronzio and W. Furmanski, {\it ibid.} {\bf B207},
1 (1981);
N. Paver and D. Treleani, {\it Nuovo Cimento} {\bf A70},
215 (1982); Zeit.
Phys.{\bf C28}, 187 (1985); B. Humpert, {\it Phys. Lett.}
{\bf 131B}, 461 (1983);
B. Humpert and R. Odorico, {\it ibid} {\bf 154B}, 211 (1985);
T. Sjostrand and
M. Van Zijl, {\it Phys. Rev.} {\bf D36}, 2019 (1987).
\item{12.} G. Calucci and D. Treleani, {\it Phys. Rev.}
{\bf D41}, 3367 (1990).
\item{13.} D. Treleani,
preprint INFN/AE-94/26, to be published on the {\it Int. J. Mod. Phys.}
{\bf A}.
\item{14.} R. Ragazzon and D. Treleani, {\it Z. Phys.}
{\bf C65}, 305 (1995).
\item{15.} J.L. Cardy and G.A.
Winbow, Phys. Lett. {\bf 52B}, 95 (1974); C.E. De Tar, S.D. Ellis and P.V.
Landshoff, Nucl. Phys. {\bf B87}, 176 (1975);
L. Bertocchi and D. Treleani, {\it J. Phys. G: Nucl. Phys.} {\bf 3}, 147
(1977).
\item{16.} F. Abe et al., {\it Phys. Rev.} {\bf D47}, 4857 (1993).
\item{17.} P.N. Harriman, A.D. Martin, W.J. Stirling and R.G. Roberts,
{\it Phys. Rev.} {\bf D42}, 798 (1990).
\vfill
\eject
\par    {\bf Figure captions}
\vskip .15in
\item{Fig. 1:} Cross section for production of minijets with $k_m\ge5GeV$.
Experimental data from UA1[2]. Dashed curves: single BFKL Pomeron
exchange with $\alpha_s=.34$ (upper curve) and $\alpha_s=.29$ (lower
curve). Continuous curves: unitarized cross section, Eq.(28) in the text,
same values of $\alpha_s$ as in the previous case and $\sigma_{eff}=20mb$.
\vskip.1in
\item{Fig. 2:} $\phi$ distribution with unitarity corrections included
(solid line) and in the single collision approximation (dashed line).
$N(\phi )$ is proportional to the differential cross section given by
eq. (27), with minijet rapidities kept fixed at $y_a=2.5$ and $y_b=-2.5$.
The normalization is such that $N(0)=N(2\pi)=1$.
\vskip.1in
\item{Fig. 3:} $\phi$ distribution for several choices of the cutoff:
$k_{min}=7GeV$ (solid line), $k_{min}=6GeV$ (dashed line), $k_{min}=5GeV$
(dotted line). $N(\phi)$ is defined as in Fig. (2) and unitarity corrections
are included.
\vskip.1in
\item{Fig.4:}
$\phi$ distribution for different choices of the rapidity gap. The cutoff
is $k_{min}=7GeV$ and the minijet rapidities are fixed at $y_{a,b}=\pm 2.5$
(solid line) and $y_{a,b}=\pm 3.5$ (dashed line).

\vfill
\eject
\bye